# Lines, dots and spirals on Peruvian land

Amelia Carolina Sparavigna
Dipartimento di Fisica,
Politecnico di Torino, Torino, Italy

The most famous geoglyphs of Peru are the "Nazca Lines". Considered as one of the mysteries of the ancient world, they have been included among the UNESCO World Heritage Sites. Located in a large region between the towns of Nazca and Palpa, these lines create shapes of animals ranging in size up to 300 m. The archaeological site is under investigation with remote science technologies.

Peru has several dry regions hosting unique remains of the ancient civilizations flourished in this country before the arrival of Spanish people. These remains are geoglyphs, which are huge drawings created on the ground. On the dry pampas (pampa is a Quechua word meaning "plain"), the drawings were made by removing the uppermost surface, exposing the underlying ground which has a different color. This technique produces a "negative" geoglyph. A "positive" geoglyphs is instead created by the arrangement of stones, gravel or earth. Peru has some positive geoglyphs near the Titicaca Lake, formed by earthworks used for cultivation [1]. Geoglyphs have been also found in some deforested regions of Amazonia [2].

The most famous negative geoglyphs of Peru are the "Nazca Lines". These Lines are considered as one of the mysteries of the ancient world and many interpretations have been proposed on their function. Included among the UNESCO World Heritage Sites in 1994, the Lines are located in the Nazca Desert, a large region between the towns of Nazca and Palpa. The figures range in size up to 300 m across, such as in Fig.1, where we can see a Frigatebird having a very long beak: because of their size, they can better recognize from the air or in satellite imagery. Sometimes it is told that the lines cannot be seen from the ground: this is not true, because these lines, such as other giant geoglyphs, can be observed from the surrounding hills. The discovery for the Western archaeology of them is credited to the Peruvian archaeologist Toribio Mejia Xesspe, who saw them in 1927 exactly during an excursion on the hills [3]. The Lines became popular when commercial airlines began flying across the Nazca desert [4]. Let us also remember Maria Reiche, the scientist that devoted a large part of her life in studying the site, starting from 1940. She gained the government recognition for the preservation of this archaeological area.

It is generally believed that the Lines were created by the Nazca culture between 400 and 650 CE. In [5], the authors reported the use of optically stimulated luminescence for dating of quartz buried when the lines were constructed, concluding that the stone lines at sites of San Ignacio and Sacramento were constructed during that period. They suggest that the lines were made in the later part of the Early Intermediate Period by people of the Nazca Culture. Made by removing the reddish pebbles covering the ground, many geoglyphs have simple geometric shapes, but more than seventy are stylized designs of animals, birds, monkeys or human figures (see the Hummingbird in Fig.2 and the Monkey in Fig.3). According to Ref.4, there are several interpretations on the purpose of these geoglyphs, but a religious intent is generally attributed to them. The simple geometric lines could be connected with the flowing of water. Some animals could be symbols of fertility. But, as told in [4], some studies are proposing that the Lines were giant astronomical calendars [6] or irrigation schemes. In fact, the Nazca Lines have been linked to the system of Puquios, which are old systems of aqueducts near Nazca. Most of them are still working and bringing fresh water to this arid land [7]. The assertion, made by David Johnson, is the following, that the Nazca Lines depict maps and pointers to the subterranean aquifers feeding the Puquios system [3,4,8]. That is, "if people knew the location of water in such a dangerously hot place, surely they'd mark it somehow" [3].

A beautiful Web site [9] shows a large collection of the geoglyphs at Nazca and nearby locations. This web site is proposing a classification of the Nazca, Palpa and Ingenio symbols according to their geometric shapes (lines, trapezoids, radiant points, spirals) and according to what geoglyphs are representing (humans, animals, plants). We can easily see, using the Google Maps and processing them as previously proposed [10], that many lines, very long, are running throughout the Nazca region, with lengths of several kilometres (see Fig.4). The lines can be narrow, with a width of a few meters, but also quite wide. There are parallel lines and trapezoids: in Ref.9, it is supposed that from the narrow side of a trapezoid, it might be possible to see a part of the horizon, where the sun and other celestial bodies rise or set during a specific period of the year. Then, the lines could be astronomical sight lines. There is also another interpretation: the lines could have been used as ceremonial avenues [4,11]. In 1985, the archaeologist Johan Reinhard proposed that the worship of mountains and water sources was the predominant feature in the Nazca religion and economy. The scientist proposed the lines and figures being part of some worship practices of deities associated with the water, to invoke their favour. .

Besides straight lines and symbols we see radiant points that are central points from which many lines radiate, zigzag lines and spirals. The lines that form zigzag patterns are usually associated with another shapes or symbols [3]. I would like to note that besides the mentioned figures, we can see arrays of "dots". If we look at Figure 5, we find an interesting geoglyph, having the shape of a line, partially covered inside with small structures that in the image of Google Maps look like dots. Probably, they are holes as in some geoglyphs of the Pisco Valley [9]. Another fact, which is in my opinion quite remarkable, is that the geoglyphs were created on the hills too. Figure 6 shows some examples: the lower panel is reporting a complex of lines with a spiral on a hill near Estudiantes and El Ingenio. Old symbol, sometimes accompanying zig-zag lines, the spiral appears as the tail of the Monkey in Fig.3.

For what concerns the tools that the Nazca people used to create the lines, Wikipedia is telling that some wooden stakes have been found at the end of some lines and that a researcher, Joe Nickell of the University of Kentucky, reproduced some figures by using the tools available to the Nazca people, confirming that an aerial assistance was not necessary to have large and complex figures. But, for new researches to be performed in this archaeological region, the remote sensing technologies are fundamental. It seems that in 2010, the Airship Nazca Project started a detailed survey of the site, using an airship equipped for the purpose [12]. According to the Web site [12], this archaeological research is interdisciplinary with geophysics. Moreover, due to the delicate texture of the Nazca lines, "the main application of geophysics will be via a range of remote sensing technologies", such photography, photogrammetry, Lidar survey, radiometry, and several other methods. It seems that the use of Lidar (Light Detection and Ranging) is particularly suitable and promising on a terrain having poorly distinguishable three-dimensional features.

**References**
[1] A. C. Sparavigna, The geoglyphs of Titicaca, Archaeogate, October 13, 2010, http://www.archaeogate.org/classica/article/1305/1/; Symbolic landforms created by ancient earthworks near Lake Titicaca, 2010, http://arxiv.org/abs/1009.2231
[2] A. C. Sparavigna, Lines under the forest, 2011, http://arxiv.org/abs/1105.5277; Archaeogate, May 30, 2011, http://www.archaeogate.org/classica/article/1411/1/
[3] Jack McClintock, The Nasca Lines Solution, Demystifying South America's gigantic archaeological puzzle, Discover Magazine, 2000, page 76.
[4] Wikipedia, http://en.wikipedia.org/wiki/Nazca_Lines
[5] W. J. Rink and J. Bartoll, Dating the geometric Nasca lines in the Peruvian desert, Antiquity, 79, 390 (2005).
[6] C. Stokes Brown, 2007, Big History, New York, The New Press, ISBN 978-1-59588-196-9

[7] K. J. Schreiber and J. Lancho Rojas, The Puquios of Nasca, Latin American Antiquity, 6, 229 (1995).
[8] D. Johnson, The Water Lines of Nazca, http://www.rumbosonline.com/articles/11-50-nazca.htm
[9] The Mystery of the Nazca Lines, http://www.nazcamystery.com/nazca_lines.htm
[10] A. C. Sparavigna, Enhancing the Google imagery using a wavelet filter, 2010, http://arxiv.org/abs/1009.1590
[11] J. Reinhard, 1996, The Nazca Lines: A New Perspective on their Origin and Meaning, Lima, Los Pinos, ISBN 84-89291-17-9
[12] The Airship Nazca Project, http://www.airshipnazca.com/#

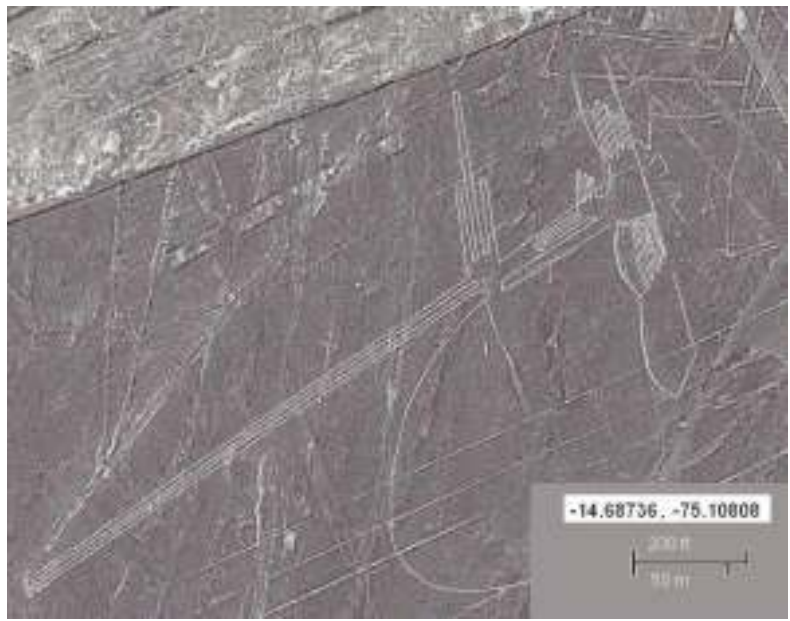

Fig.1 Two geoglyphs of Nazca represent a Frigatebird and a Heron. They are negative geoglyphs created by removing the uppermost surface and exposing the underlying ground which has a different color. The image was obtained after processing a Google Maps image and marking the lines with white pixels. Note that the beak of the Frigatebird is approximately 300 meters long, and composed by perfect straight lines.

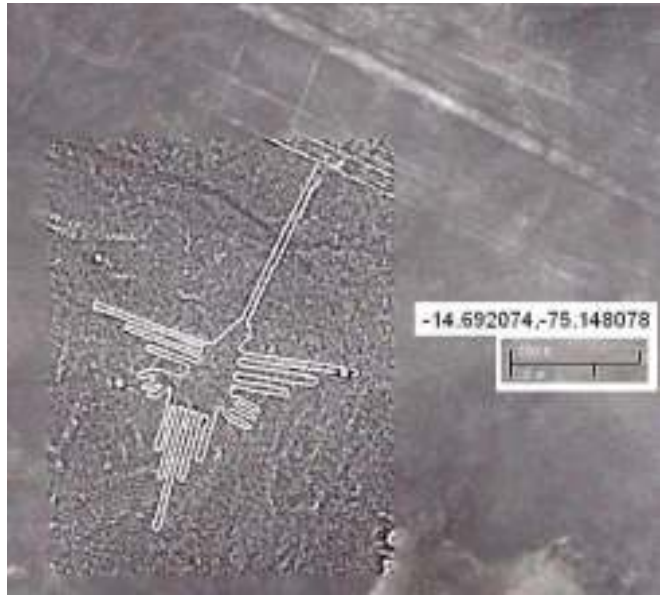

Fig.2 Here the Hummingbird. The image was obtained after processing a Google Maps image and marking the drawing with white lines.

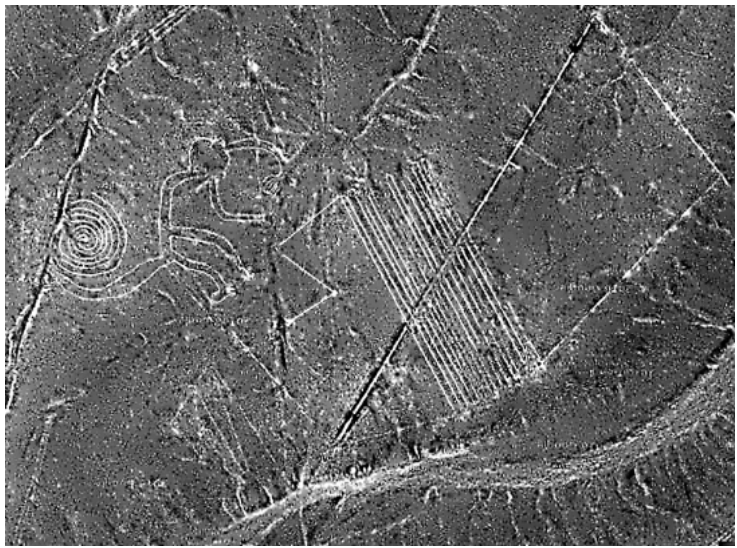

Fig.3 The geoglyph of the Monkey (after processing a Google Maps image).

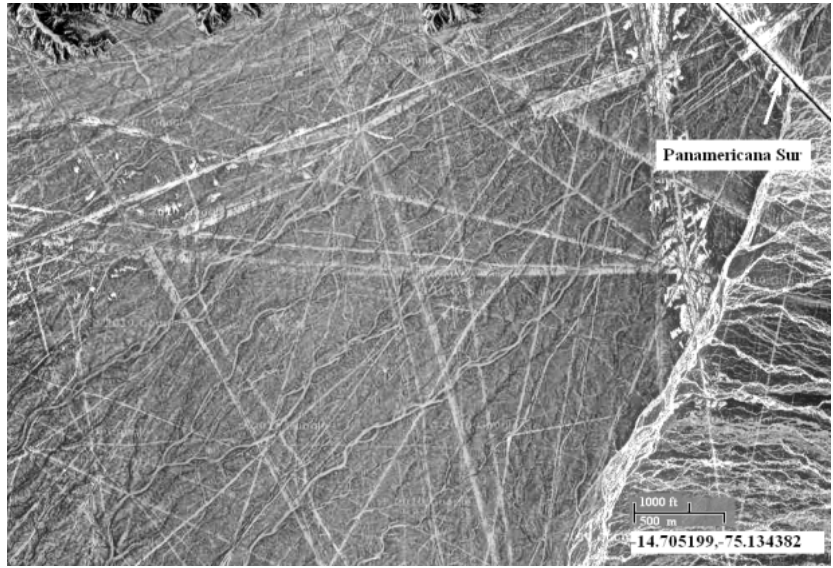

Fig.4 We can easily see, using the Google Maps and processing them, that many very long lines are running throughout the Nazca region, with lengths of several kilometres. The lines can be narrow or very wide. There are also parallel lines and trapezoids. Some lines could be astronomical sight lines or ceremonial avenues [9,11].

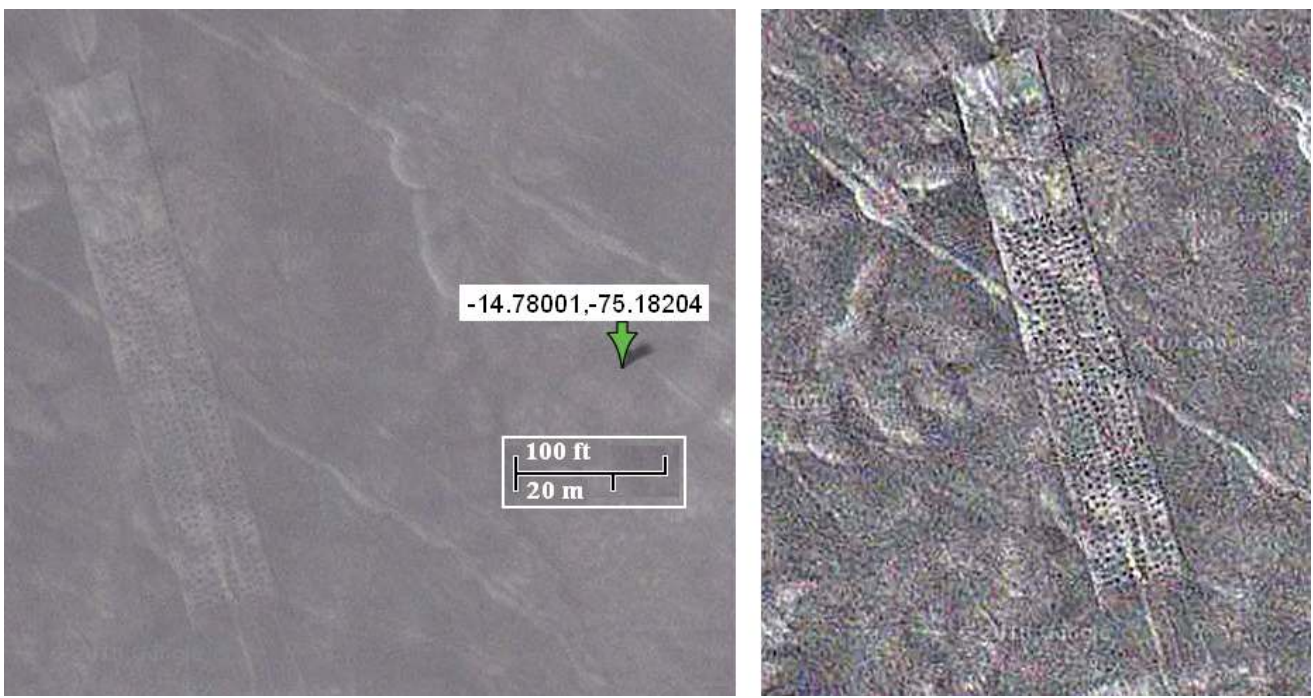

Fig.5 Not only lines. This is an interesting geoglyph, having the shape of a line, but it is covered inside with small structures that, in the satellite image of Google Maps, look like dots.

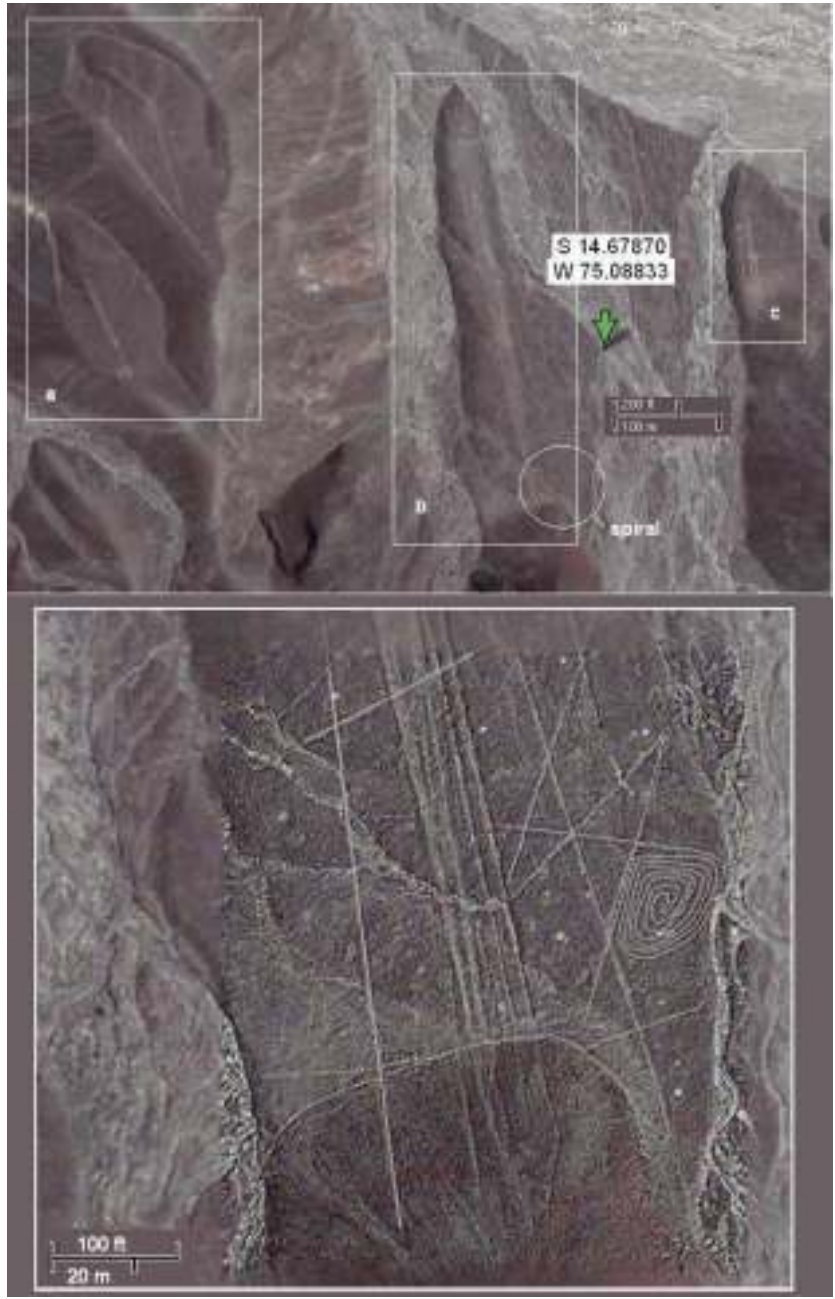

Fig.6 These images show that we can see geoglyphs not only on the plains. Several of them are on the hills. The lower panel reproduces a complex of lines with a spiral.